\documentclass[aps,pre,twocolumn,groupedaddress,showpacs,floatfix]{revtex4}
\usepackage{dcolumn}
\usepackage{amsmath}
\usepackage{amssymb}
\usepackage{graphicx}
\usepackage{bm}
\usepackage[T1]{fontenc}
\usepackage{color}
\usepackage{url}
\usepackage{rotating}

\begin{document}

\title{Effect of assortative mixing in the second-order Kuramoto model}

\author{Thomas K. DM. Peron$^1$}
\email{thomas.peron@usp.br}
\author{Peng Ji$^{2,3}$}
\email{pengji@pik-potsdam.de}
\author{Francisco A. Rodrigues$^4$}
\email{francisco@icmc.usp.br}
\author{J\"urgen Kurths$^{2,3,5}$}
\affiliation{$^1$Instituto de F\'{\i}sica de S\~{a}o Carlos, Universidade de S\~{a}o Paulo, S\~{a}o Carlos, S\~ao Paulo, Brazil\\
$^2$Potsdam Institute for Climate Impact Research (PIK), 14473 Potsdam, Germany\\
$^3$Department of Physics, Humboldt University, 12489 Berlin, Germany\\
$^4$Departamento de Matem\'{a}tica Aplicada e Estat\'{i}stica, Instituto de Ci\^{e}ncias Matem\'{a}ticas e de Computa\c{c}\~{a}o,Universidade de S\~{a}o Paulo, Caixa Postal 668,13560-970 S\~{a}o Carlos,  S\~ao Paulo, Brazil\\
$^5$Institute for Complex Systems and Mathematical Biology, University of Aberdeen, Aberdeen AB24 3UE, United Kingdom}

\begin{abstract}
In this paper we analyze the second-order Kuramoto model presenting a positive correlation between the heterogeneity of the connections and the natural frequencies in scale-free networks. We numerically show that discontinuous transitions emerge not just in disassortative but also in assortative networks, in contrast with the first-order model. We also find that the effect of assortativity on network synchronization can be compensated by adjusting the phase damping. Our results show that it is possible to control collective behavior of damped Kuramoto oscillators by tuning the network structure or by adjusting the dissipation related to the phases movement.

\end{abstract}


\maketitle

\section{Introduction}

Synchronization is pervasive in nature, society and technology~\cite{Arenas08:PR}. This collective behavior emerges from the interaction of neurons in the central nervous system, power grids, crickets, heart cells and lasers~\cite{Pikovsky03, Arenas08:PR}. Synchronization arises due to the adjustment of rhythms of self-sustained periodic oscillators weakly connected~\cite{Pikovsky03, Acebron05:RMP, Arenas08:PR} and can be treated mathematically by the model proposed by Kuramoto~\cite{Acebron05:RMP}. The general Kuramoto model assumes that the natural frequencies of the oscillators are selected from unimodal and symmetric random distributions~\cite{Arenas08:PR}. In this case, a second-order phase transition to synchronization can be observed~\cite{Acebron05:RMP, Arenas08:PR}. However, the first-order Kuramoto model can exhibit discontinuous phase transitions~\cite{pazo2005thermodynamic,basnarkov2007phase,basnarkov2008kuramoto}. For instance, in one of the first works on this topic, Paz\'o~\cite{pazo2005thermodynamic} showed that, if uniform frequency distributions are considered,
first-order transitions emerge in fully connected Kuramoto oscillators. Very recently,   G\'omez-Garde\~nes \textit{\textit{et al.}}~\cite{Gardenes011:PRL} verified that a discontinuous synchronization transition can also occur  in scale-free networks as an effect of positive correlation between the natural frequencies and network topology. This discovery has triggered many ensuing works, which analyzed explosive synchronization analytically and numerically~\cite{peron2012determination,Leyva012, coutinho2013kuramoto,skardal2013effects, chen2013explosive, su2013explosive, zhang2013explosive,zou2014basin,zhang2015explosive}.

The early works on explosive synchronization (e.g.~\cite{peron2012determination, Leyva012, coutinho2013kuramoto}) suggested that the correlation between frequency and degree distributions is the only condition required for the emergence of a discontinuous synchronization transition in scale-free networks. However, subsequent papers have shown that different criteria to set the frequency mismatch between the oscillators~\cite{zhu2013criterion}, the presence of time-delay~\cite{peron2012explosive}, non-vanishing degree-degree correlation~\cite{li2013reexamination, zhu2013criterion, liu2013effects} or the inclusion of noise~\cite{sonnenschein2013networks} can dramatically change the type of the phase transitions, even in the regime of fully connected graphs~\cite{zhu2013criterion,leyva2013explosive}. 

In the case of degree-degree correlation, Li \textit{\textit{et al.}}~\cite{li2013reexamination} verified that assortative scale-free networks no longer undergo a discontinuous transition, even if the network presents a positive correlation between structural and dynamical properties. This behavior was also observed in the synchronization of FitzHugh-Nagumo (FHN) oscillators coupled in scale-free networks under the constraint of correlating frequencies and degrees~\cite{chen2013explosive}. Furthermore, Zhu \textit{et al.}~\cite{zhu2013criterion} found that discontinuous transitions only emerge in networks subjected simultaneously to negative degree-degree and frequency-frequency correlations. Effects of degree-degree correlation on general network synchronization phenomena were also analyzed in literature (cf. \cite{chavez2006degree,di2007effects,sorrentino2007synchronizability} for studies in the context of the master stability function formalism). For instance, Bernardo \textit{et al.}~\cite{di2007effects} studied scale-free networks of identical R\"ossler oscillators and showed that disassortative mixing enhances network synchronization, when compared with uncorrelated networks~\cite{di2007effects}. On the other hand, regarding the synchronization of weighted networks, assortative mixing can enhance synchronization, depending on the weighting procedure~\cite{chavez2006degree}.

All these works considered only the first-order Kuramoto model. However, in various systems, the second-order Kuramoto model is more suitable to describe the emergence of synchronization~\cite{strogatz1994nonlinear,Tanaka1997279}. Indeed, many real-world systems, such as power-grid networks~\cite{filatrella2008analysis,
dorfler2012synchronization,dorfler2013synchronization,rohden2012self,menck2013basin,menck2014dead,rohden2014impact,doerfler2014survey},     superconducting Josephson junctions~\cite{strogatz1994nonlinear} and many other applications~\cite{strogatz1994nonlinear,Tanaka1997279}, can be modeled as networks whose units are second-order Kuramoto oscillators. 

In the context of explosive synchronization, the second-order model was introduced only recently~\cite{PhysRevLett.110.218701,ji2014analysis}. In contrast to~\cite{Gardenes011:PRL}, where the authors verified that nodes in scale-free network join the synchronous group abruptly giving rise to a discontinuous synchronization transition, we have shown that this behavior is no longer observed when an inertia term is included~\cite{PhysRevLett.110.218701,ji2014analysis}. Despite also observing a discontinuous transition of the order parameter, we verified that in the second-order Kuramoto model with frequencies proportional to degrees, nodes join the synchronous component successively grouped into cluster of nodes with the same degree, a phenomenon called \textit{cluster explosive synchronization} (CES)~\cite{PhysRevLett.110.218701}.

The influence of network structure on the emergence of explosive synchronization in the second-order Kuramoto model proposed in~\cite{PhysRevLett.110.218701} has not been addressed yet, since only uncorrelated networks have been considered~\cite{PhysRevLett.110.218701, ji2014analysis}. Among important network properties, the degree-degree correlation is observed in several complex networks~\cite{newman2002assortative,newman2003mixing, Costa011:AP} and 
it plays a fundamental role in many dynamical processes, such as epidemic spreading and synchronization~\cite{Barrat08:book}. For example, the degree-degree correlation can change the type of phase transitions of the first-order Kuramoto model with a positive correlation between frequency and degree distributions~\cite{ li2013reexamination, zhu2013criterion}. In this way, since the frequency mismatch between oscillators has been shown to play a crucial role in the emergence of abrupt transitions in the first-order Kuramoto model, it is natural to ask about the effects of degree-degree correlations on the overall dynamics in models with inertia. 

In this paper we study the second-order Kuramoto model in networks with degree-degree correlations, i.e., non-vanishing assortativity. We find that the synchronization diagrams have a strong dependence on the network assortativity, but in a different fashion compared to the first-order model~\cite{li2013reexamination}. In fact, for negative and positive assortativity values, the synchronization is observed to be discontinuous, depending on the damping coefficient. Moreover, the upper branch in the synchronization diagrams associated to the case in which the coupling is decreased is barely affected by different assortativity values, again in contrast with   the first-order Kuramoto model~\cite{chen2013explosive,li2013reexamination,zhu2013criterion,zhang2013explosive,su2013explosive,liu2013effects}. 
In other words, we show here that one is able to
control the hysteretic behaviour of the second-order Kuramoto model by tuning the network properties, the phenomenon that was not investigated
before.
In order to compare with different choices of frequencies distributions, we also investigate the dynamics of damped Kuramoto oscillators in assortative networks using unimodal and even distributions, without being correlated with the local topology. Similarly as in the case of frequencies proportional to degrees, we again observe very similar behavior for the onset of synchronization over networks with different degree-degree correlations. 

\section{Synchronization in correlated networks}

We consider networks where each node is a phase oscillator evolving according to the second-order Kuramoto model~\cite{Acebron05:RMP,Arenas08:PR}
\begin{equation}
\frac{d^2\theta_i}{dt^2}=-\alpha \frac{d\theta_{i}}{dt} + \Omega_{i} + \lambda \sum_{j=1}^{N}A_{ij} \sin\left[\theta_{j}(t) - \theta_{i}(t) \right],
\label{eq:kuramoto} 
\end{equation}
where $\alpha$ is the dissipating parameter, $\lambda$ is the coupling strength and $\Omega_{i}$ is the natural frequency of oscillator $i$, defined according to a given probability distribution $g(\Omega)$. The heterogeneity of the network connections is accounted by the adjacency matrix $\mathbf{A}=\left\{A_{ij} \right\}$, whose elements $A_{ij}=1$ if oscillators $i$ and $j$ are connected, and $A_{ij}=0$ otherwise. The collective dynamics of the oscillators is measured by the macroscopic order parameter, defined as
\begin{equation}
r(t)e^{i\psi(t)} = \frac{1}{N}\sum_{j=1}^{N}e^{i\theta_{j}(t)},
\end{equation}
where the modulus $0 \leq r(t) \leq 1$ and $\psi(t)$ is the average phase of the oscillators. The system governed by Eqs.~\ref{eq:kuramoto} exhibits hysteretic synchrony~\cite{PhysRevLett.78.2104,Tanaka1997279}. The onset of synchronization ($r>0$) is characterized by a critical coupling $\lambda_c^{\textrm{I}}$ when the coupling strength is progressively increased from a given $\lambda_0$. On the other hand, starting at synchronouos state and decreasing progressively the coupling strength, the oscillators fall into an incoherent state ($r \approx 0$) at coupling $\lambda_c^{\textrm{D}} \leq \lambda_c^{\textrm{I}}$~\cite{PhysRevLett.78.2104,Tanaka1997279}.

Here we study the second-order Kuramoto model (see Eq.~\ref{eq:kuramoto}) in which the natural frequency distribution $g(\Omega)$ is correlated with the degree distribution $P(k)$ as $\Omega_{i}=k_{i}-\left\langle k \right\rangle$~\cite{PhysRevLett.110.218701}, where $k_{i}$ is the degree of the oscillator $i$ and $\left\langle k \right\rangle$ is the average degree of the network. At first glance that particular choice for the frequency assignment could sound odd, however it is not difficult to find physical scenarios where this configuration is plausible. For example, such correlation between dynamics and network topology can arise as a consequence of a limited amount of resources or energy supply for the oscillators. In fact, studies on optimization of synchronization in complex networks~\cite{brede2008synchrony,buzna2009synchronization,skardal2014optimal} have shown that,  for a given fixed set of allowed frequencies $\{\Omega_1,\Omega_2,...,\Omega_N\}$, the configuration that maximizes the network synchronization is reached for cases in which frequencies are positively correlated
with degrees. Therefore, this correlation between frequencies and local topology can be seen as an optimal scenario for the emergence of collective behavior in complex networks. 

\begin{figure}[!t]
\includegraphics[width=\linewidth]{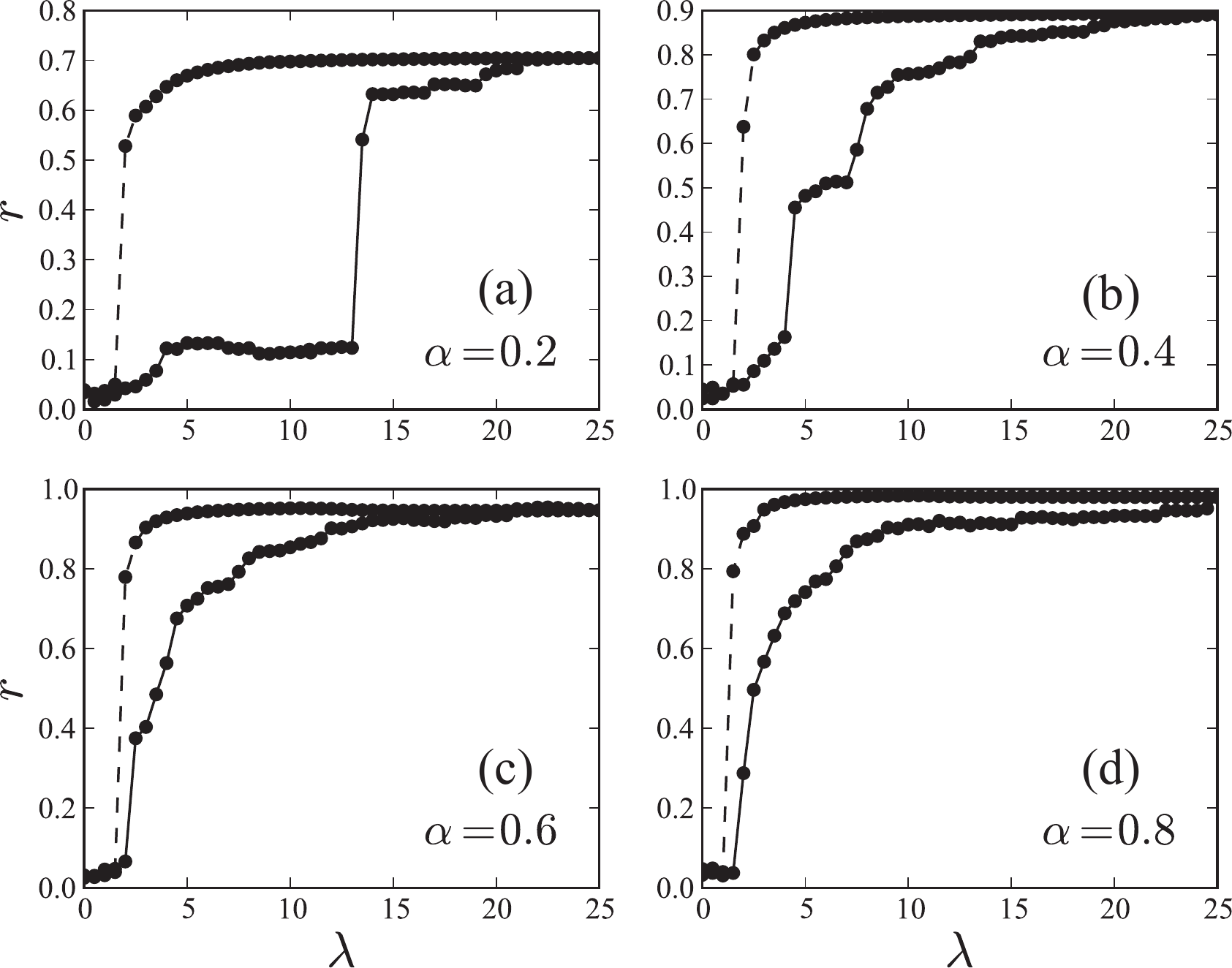}
\caption{Synchronization diagram $r(\lambda)$ with (a) $\alpha=0.2$, (b) $\alpha=0.4$, (c) $\alpha=0.6$  and (d) $\alpha=0.8$ for assortativity  $\mathcal{A}=-0.3$. With increasing $\alpha$, onset of synchronization and hysteresis decreases. The natural frequency of each oscillator is $\Omega_i = k_i - \left\langle k \right\rangle$ and the networks have $N=10^3$ and $\left\langle k \right\rangle = 6$. The degree distribution follows a power-law $P(k) \sim k^{-\gamma}$, where $\gamma=3$. Curves in which points are connected by solid lines (resp. dashed lines) correspond to the forward (resp. backward) continuations of the coupling strength $\lambda$. }
\label{fig1}
\end{figure}

We study networks presenting non-vanishing degree-degree correlation. Such a correlation is quantified by a measure known as assortativity coefficient, $\mathcal{A}$ , which is the Pearson coefficient between degrees at the end of each link~\cite{newman2002assortative}, i.e.
\begin{equation}
\mathcal{A}=\frac{M^{-1}\sum_{i}j_{i}k_{i}-\left[M^{-1}\sum_{i}\frac{1}{2}(j_{i}+k_{i})\right]^{2}}{M^{-1}\sum_{i}\frac{1}{2}(j_{i}^{2}+k_{i}^{2})-\left[M^{-1}\sum_{i}\frac{1}{2}(j_{i}+k_{i})\right]^{2}},
\label{eq:assortativity}
\end{equation}
where $-1 \leq \mathcal{A} \leq 1$,  $j_i$ and $k_i$ are the degrees associated to the two ends of the edge $i$ ($i=1,...,M$) and $M$ is the total number of edges in the network. In order to tune the degree of assortativity of each network, we use the method proposed in~\cite{xulvi2004reshuffling}. The algorithm allows us to obtain networks with a desired value of assortativity without changing the degree of each node. At each step, two edges are selected at random and the four nodes associated to these edges are ordered from the lowest to the highest degree. In order to produce assortative mixing ($\mathcal{A}>0$), with a probability $p$, one new edge connects the first and the second node and another new edge links the third and fourth nodes. In the case when one of the two new edges already exists, the step is discarded and a new pair of edges are chosen. This same heuristic can also generate disassortative networks ($\mathcal{A}<0$) with only a slight change in the algorithm. After selecting the four nodes and sorting them with respect to their degrees, one must rewire, with probability $p$, the highest degree node with the lowest one and, likewise, the second and third nodes. After rewiring the network, if the degree of assortativity is higher or smaller than the designed $\mathcal{A}$, $p$ is decreased or increased respectively and the network is rewired following the procedures described above. In order to avoid dead loops, the increasing and decreasing steps of $p$ should not be equally spaced.

\section{Numerical Results}

In this section we present the results obtained by numerically evolving the equations of motions considering Eq.~\ref{eq:kuramoto} in assortative networks constructed according to the model described in the previous section. In all simulations the initial networks are constructed through the Barab\'asi-Albert (BA) model with $\left\langle k \right\rangle = 6$ and $N=1\times 10^3$.

The order parameter $r$ is calculated with forward and backward continuations of the coupling strength $\lambda$. More specifically, by increasing the value of $\lambda$ adiabatically, we integrate the system long enough and calculate the stationary value of $r$ for each coupling $\lambda_{0}, \lambda_{0}+\delta \lambda, ..., \lambda_{0}+n\delta \lambda$. Similarly, for the backward continuation, we start at the value $\lambda = n \delta \lambda + \lambda_{0}$ and decrease $\lambda$ by amounts of $\delta \lambda$ until $\lambda=\lambda_{0}$. In both processes we use $\delta \lambda=0.5$. 

\begin{figure}[!t]
\includegraphics[width=\linewidth]{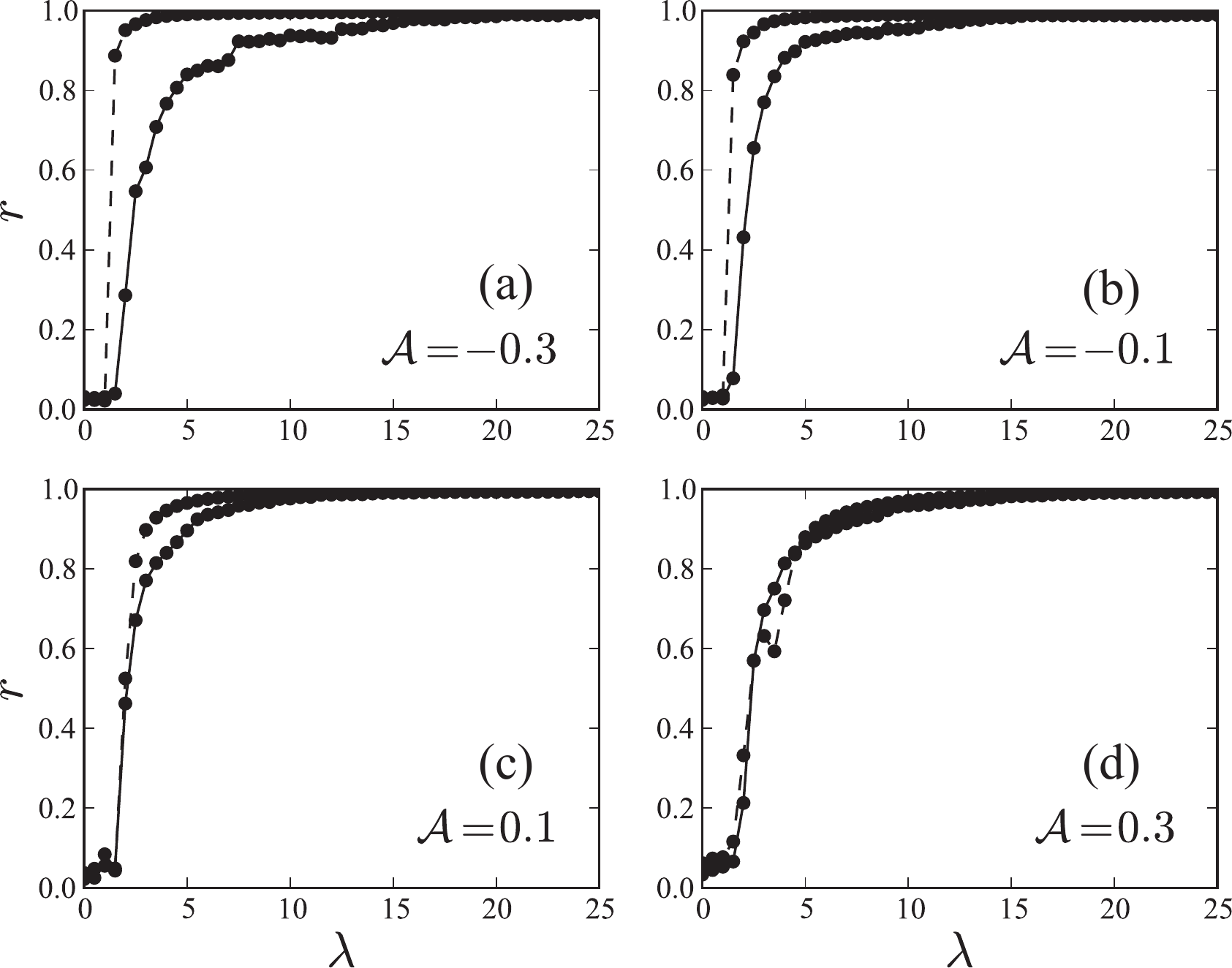}
  \caption{Synchronization diagram $r(\lambda)$ for (a) $\mathcal{A}=-0.3$, (b) $\mathcal{A}=-0.1$, (c) $\mathcal{A}=0.1$ and (d) $\mathcal{A}=0.3$. The dissipation coefficient is fixed at $\alpha=1$ with natural frequencies given by $\Omega_i = k_i - \left\langle k \right\rangle$, as in Fig.~\ref{fig1}. All networks considered have $N=10^3$, $\left\langle k \right\rangle = 6$ and $P(k) \sim k^{-\gamma}$, where $\gamma=3$.}
  \label{fig2}
\end{figure}

 \begin{figure*}[htp!]
\includegraphics[width=1.0\linewidth]{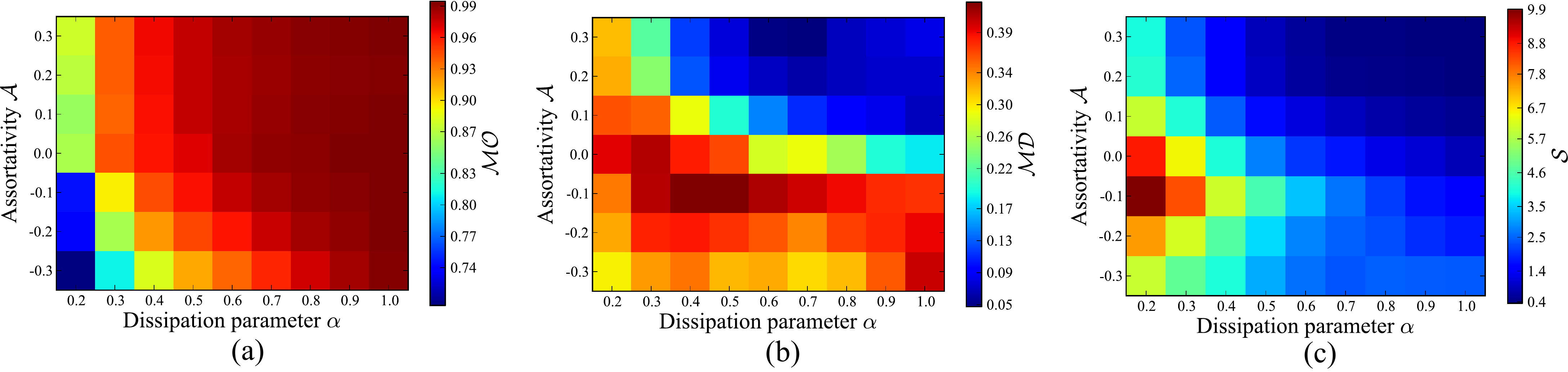}
\caption{ (Color online) Contour plot on $\alpha$-$\mathcal{A}$ plane colored according to (a) the maximal order parameter $\mathcal{MO}$, (b) the maximal order parameter difference $\mathcal{MD}$ and (c) the hysteresis area $\mathcal{S}$. $\mathcal{A}$ and $\alpha$ are varied within the interval $[-0.3,0.3]$ and $[0.2,1]$, respectively. For each pair $(\alpha,\mathcal{A})$, $40$ times simulations are performed with the coupling strength in the interval as in Fig.~\ref{fig1}, i.e, $\lambda \in [0,25]$.
}
\label{fig3}
\end{figure*}
  
We investigate the dependency of the hysteresis on the dissipation parameter $\alpha$. Fig.~\ref{fig1} shows the forward and backward synchronization diagrams $r(\lambda)$ for networks with assortativity $\mathcal{A}=-0.3$, but different values of $\alpha$ within the interval $[0.2,1]$. As we can see, the area of hysteresis and the critical coupling for the onset of synchronization in the increasing branch tends to decrease as $\alpha$ is increased,
which also contributes to increase the maximal value of the order parameter. 

Next, we fix the dissipation coefficient $\alpha=1$ and vary the network assortativity in the interval $[-0.3,0.3]$.  Fig.~\ref{fig2} shows the synchronization diagram $r(\lambda)$  for networks with different values of assortativity. As $\mathcal{A}$ increases, the hysteresis becomes less clear and the onset of synchronization in the decreasing branch tends to increase. Surprisingly, the critical coupling of the increasing branch for the second-order Kuramoto model is weakly affected, which is in sharp contrast with results concerning models without inertia~\cite{li2013reexamination, liu2013effects}. More precisely, in the first-order Kuramoto model with frequencies correlated with degrees, the critical coupling for the onset of synchronization in scale-free networks increases as the network becomes more assortative~\cite{li2013reexamination,liu2013effects}. The same phenomenon was observed in the synchronization of FHN oscillators~\cite{chen2013explosive}. 

\begin{figure}[!b]
\includegraphics[width=\linewidth]{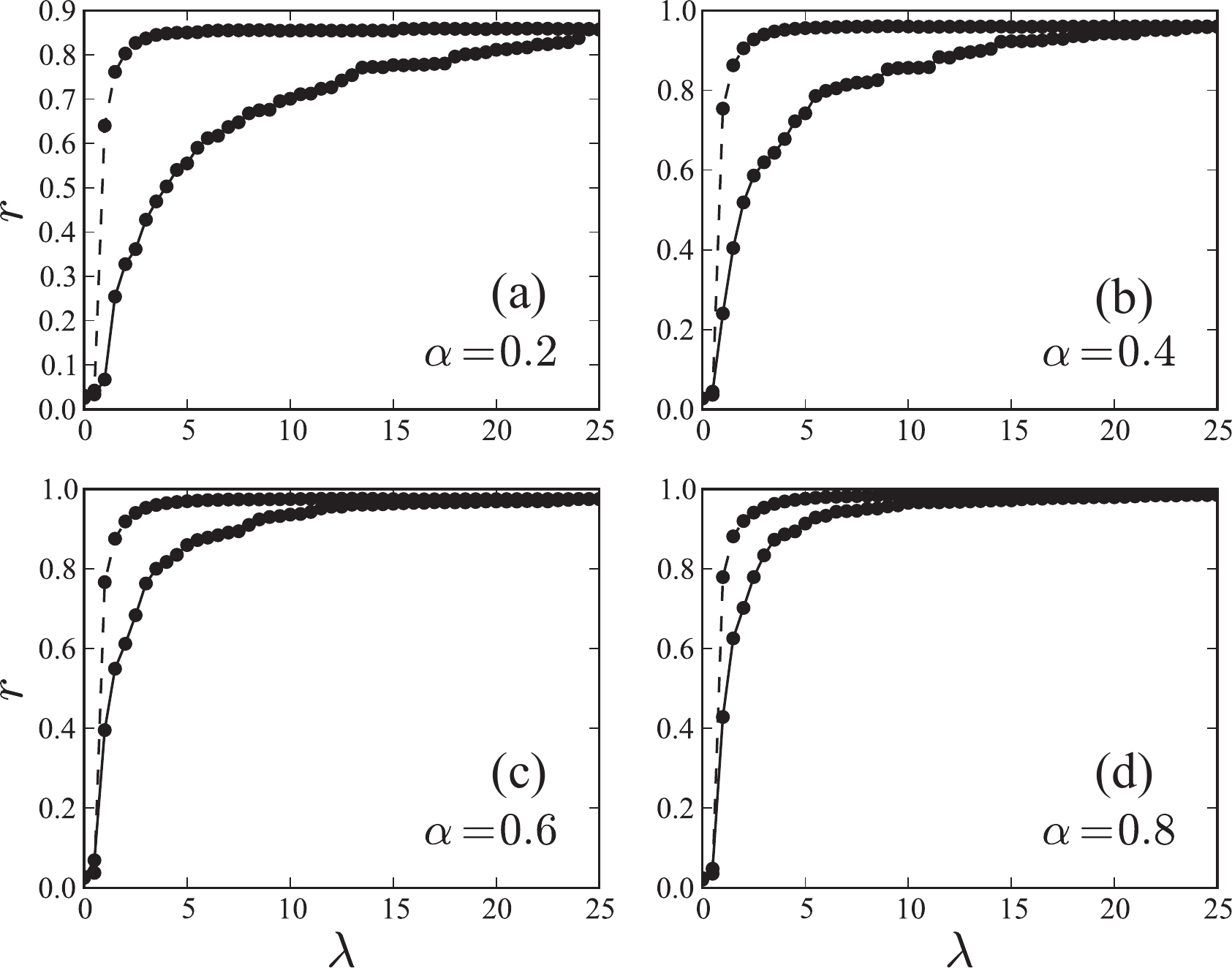}
\caption{Parameters are the same as in Fig.~\ref{fig1}, except that the natural frequencies are randomly selected from a Lorentzian distribution $g(\Omega) = 1 / (\pi(1+\Omega^2))$ .}
\label{fig4}
\end{figure}

\begin{figure}[!b]
\includegraphics[width=1.0\linewidth]{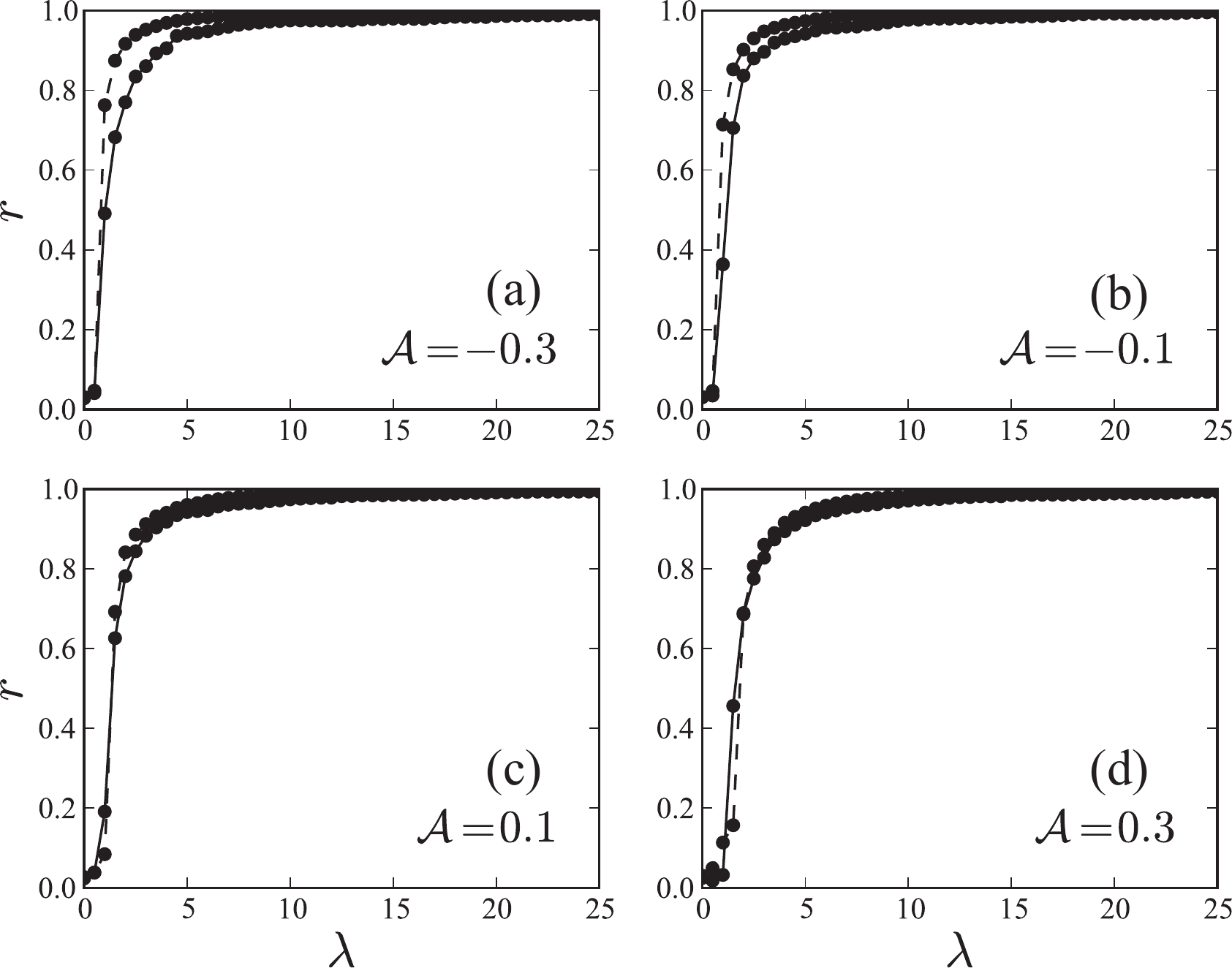}
 \caption{Parameters are the same as in Fig.~\ref{fig2}, except that natural frequencies are randomly selected from a Lorentzian distribution $g(\Omega) = 1 / (\pi(1+\Omega^2))$. }
  \label{fig5}
\end{figure}
\begin{figure*}[!htp]
\includegraphics[width=1.0\linewidth]{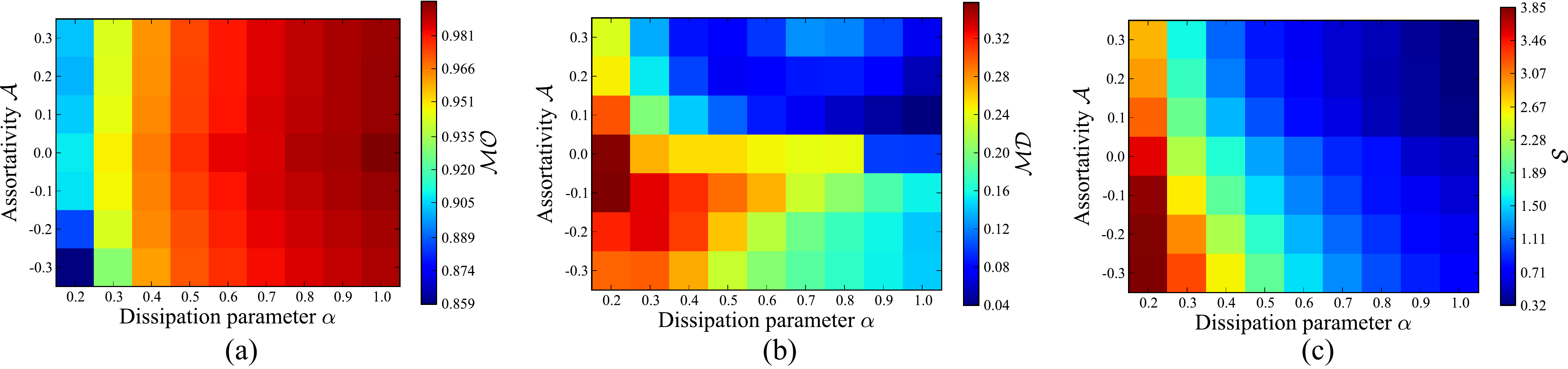}
\caption{ (Color online) Contour plots similar as in Fig.~\ref{fig3} for networks with a natural frequency distribution given by $g(\Omega) = 1/(\pi(1 + \Omega^2))$. }
\label{fig6}
\end{figure*} 
In order to evaluate more accurately the dependency of the synchronization transitions on the level of assortativity and the dissipation parameter, we introduce the maximal order parameter $\mathcal{MO}$, the maximal order parameter difference $\mathcal{MD}$ and the hysteresis area $\mathcal{S}$ in the synchronization diagrams as a function of $\mathcal{A}$ and $\alpha$, respectively, as follows, 
\begin{equation}
\mathcal{MO}=\left\langle \max{(r(\lambda))}\right\rangle,
\label{eq:MO}
\end{equation}
\begin{equation}
\mathcal{MD}=\left\langle \max{|r^{\textrm{I}}(\lambda)-r^{\textrm{D}}(\lambda)|}\right\rangle,
\label{eq:MD}
\end{equation}
\begin{equation}
\mathcal{S}=\left\langle \int |r^{\textrm{I}}(\lambda)-r^{\textrm{D}}(\lambda) |d\lambda \right\rangle,
\label{eq:S}
\end{equation}
where $\lambda \in [\lambda_0, \lambda_0+n\delta\lambda]$, $\left\langle \centerdot \right\rangle$ denotes the average of different realizations and $|\centerdot|$ the absolute value. $r^{\textrm{I}}(\lambda)$ and $r^{\textrm{D}}(\lambda)$ are the order parameters for increasing and decreasing coupling strength $\lambda$, respectively. If $r(\lambda)$ increases as $\lambda$ grows, then $\mathcal{MO}$ is usually obtained at the maximal coupling strength $\lambda_0 + n\delta\lambda$, i.e. $\mathcal{MO}=r(\lambda_0 + n\delta\lambda)$. $\mathcal{MD} \in [0,1]$ quantifies the hysteresis difference.  If the system shows a continuous phase transition with a perfect match between increasing and decreasing coupling strength diagrams, then $\mathcal{S}\simeq 0$. In this way, $S$ is a quantitative index to evaluate the hysteretic behavior related to the emergence of synchronization. 

Comparing Figs.~\ref{fig1} and~\ref{fig2} we observe a clear dependence of the synchronization diagrams on the assortativity $\mathcal{A}$ and on the dissipation parameter $\alpha$. Moreover, note that it is also possible to obtain similar dependencies of $r$ on $\lambda$ by selecting different values of $\alpha$ and $\mathcal{A}$. In order to better grasp 
this apparent equivalence in the dynamical behavior of the system for different choices of the parameters $\mathcal{A}$ and 
$\alpha$, we show in Fig.~\ref{fig3} the quantities defined in Eqs.~\ref{eq:MO},~\ref{eq:MD} and~\ref{eq:S} 
as a function of $\alpha$ and $\mathcal{A}$. As we can see in Fig.~\ref{fig3}(a), similar values for the 
maximal order parameter $\mathcal{MO}$ are obtained according to the initial setup of the model. More specifically, the level of 
synchronization of the network can be chosen by tuning the assortativity or the dissipation parameter in the dynamical model. 
Therefore, for the second-order Kuramoto model in the case of frequencies positively correlated with degree, high levels of coherent 
behavior are obtained by either strongly assortative or disassortative networks, once the dissipation parameter $\alpha$ is properly 
selected. Interestingly, the maximal gap between the increasing and decreasing branches quantified by $\mathcal{MD}$ 
(Fig.~\ref{fig3}(b)) has a maximum around $\mathcal{A} \simeq -0.1$ and $\alpha \in [0.4,0.5]$, showing 
that the area of metastability in the stability diagram of the model~\cite{PhysRevLett.110.218701} is 
maximized for this set of parameters.

A similar effect can also be observed for the hysteresis area $\mathcal{S}$ in Fig.~\ref{fig3}(c). The maximal $\mathcal{S}$ in the synchronization diagram is reached for networks with $\mathcal{A}=-0.1$ and $\alpha=0.2$. Furthermore, similar values of $\mathcal{S}$ are obtained by different sets of $\alpha$ and $\mathcal{A}$, which shows an interesting interplay between the topological parameter (assortativity) and the dynamical one (dissipation). More precisely, topological properties related to degree-degree correlations can be counterbalanced by the dissipation parameter in the dynamical model. This property could have interesting applications in the control of synchronization in networks modeled by the second-order Kuramoto model. In particular, if one is interested 
to reduce hysteresis in a system, such task can be accomplished by either increasing the dissipation or the degree mixing in the network. 
Therefore, the question usually addressed in studies regarding the first-order Kuramoto model that is whether assortativity could enhance synchronization or not~\cite{li2013reexamination,zhu2013criterion} turns out to be harder to answer for the damped version of the model. The reason for that is that the asymptotic behavior of the system strongly depends on the combination of parameters $\mathcal{A}$ and $\alpha$, which allows at the same time much more options to control the system by tuning such parameters.

In order to analyze how assortative mixing influences the dynamics of networks of damped Kuramoto oscillators
without the constraint of having $\Omega_i \propto k_i$, we also compute the same forward and backward synchronization diagrams
considering a Lorentzian distribution $g(\Omega)=\frac{1}{\pi(1+\Omega^2)}$  for different values of degree assortativity $\mathcal{A}$. Similarly as before, as a
first experiment, we fix the assortativity  at $\mathcal{A}=-0.3$ and vary the dissipation parameter $\alpha$ as indicated in 
Fig.~\ref{fig4}. Again, as we increase $\alpha$ the hysteresis area tends to decrease. The same
effect is observed for a fixed $\alpha$ with varying $\mathcal{A}$, as depicted in Fig.~\ref{fig5}.

Calculating  $\mathcal{MO}$ as a function of $\alpha$ and $\mathcal{A}$ we note, however, a slight
different dependence compared to the case where frequencies are proportional to degree. As we can see in Fig.~\ref{fig6}(a), for a fixed value of $\alpha$, $\mathcal{MO}$ is weakly affected by the change
of the degree mixing, except for the case $\alpha=0.2$. Nonetheless, the model with frequencies correlated with degrees
presents larger fluctuations for the maximum value of coherent behavior, comparing Figs.~\ref{fig3}(a) and~\ref{fig6}(a). Furthermore, the maximum value of $\mathcal{MD}$ (Fig.~\ref{fig6}(b)) for Lorentzian frequency distributions is obtained for slight disassortative networks with low values of the dissipation parameter $\alpha$. We also note that, for large $\alpha$, the maximum gap $\mathcal{MD}$ starts to decrease, in contrast to the case with frequencies correlated with degree (Fig.~\ref{fig3}(b)), where
$\mathcal{MD}$ is close to zero for almost the entire range considered of $\alpha$ for which $\mathcal{A}>0$. Finally, analyzing $\mathcal{S}$ in Fig.~\ref{fig6}(c) we note the 
same interplay between topological perturbations in the networks, accounted by changes in assortative mixing, and dynamical features in the oscillator model characterized by the dissipation parameter $\alpha$. As shown in this figure, similar values of $S$ are achieved by controlling the parameters $\alpha$ and $\mathcal{A}$, and highly and poorly hysteretic  synchronization diagrams can be obtained by different strategies, i.e., changing the network structure ($\mathcal{A}$) or the dynamical nature of the oscillators ($\alpha$).

\section{Conclusions}

First-order synchronization transitions for the Kuramoto model in complex networks have been known as a consequence of positive correlation between network structure, represented by the degree distribution, and the intrinsic oscillatory dynamics, represented by the natural frequency distribution of the oscillators~\cite{Gardenes011:PRL, peron2012determination, Leyva012, coutinho2013kuramoto}.

Here, we have numerically shown that such transitions for the second-order Kuramoto model also depend on the degree mixing in the network connection. More precisely, discontinuous transitions of networks of second-order Kuramoto oscillators can take place not only in disassortative ones but also in assortative ones, in contrast to what has been observed for the first-order Kuramoto model in which the correlation between topology and dynamics is also present~\cite{li2013reexamination}. The reason behind this phenomenon can be regarded as an effect of the dynamical equivalence of changes in the network structure, played by assortative mixing, and changes in the oscillator model (dissipation parameter). In other words, a given final configuration of a network of second-order Kuramoto oscillators can be achieved by tuning the network structure or by adjusting the dissipation related to the phases movement. As previously mentioned, this finding can have important applications on controlling network synchronization where, for instance, there are costs associated to lead the system to a given desirable state, allowing the adoption of different strategies to accomplish such task.

Our results show that the hysteretic behavior of the order parameter vanishes for some assortativities, suggesting that the transition might become continuous. However, to properly determine the nature of the phase transition, a more detailed study should be addressed. Moreover, the theoretical description of sychronization in correlated networks is still an open problem and mean- field theories that account for degree-degree correlations should also be developed. As a future study, it would be interesting to further analyze the present model relating the recent approaches on mean-field approximation of first-order Kuramoto oscillators in assortative networks~\cite{restrepo2014mean} and the low-dimensional behavior of the second-order model~\cite{ji2014low}.  

\section*{Acknowledgments}

TP would like to acknowledge FAPESP (No. 2012/22160-7) and IRTG 1740. PJ would like to acknowledge China Scholarship
Council (CSC) scholarship. FAR would like to acknowledge
CNPq (No. 305940/2010-4), FAPESP (No. 2010/19440-
2) and IRTG 1740 for the financial support given to this research. JK would like to acknowledge IRTG 1740
(DFG and FAPESP) for the sponsorship provided.

\bibliographystyle{apsrev}
\bibliography{paper}
\newpage

\end{document}